# WeTICA: A Directed Search Weighted Ensemble Based Enhanced Sampling Method to Estimate Rare Event Kinetics in a Reduced Dimensional Space

Sudipta Mitra, Ranjit Biswas and Suman Chakrabarty*

Department of Chemical and Biological Sciences, S. N. Bose National Centre for Basic Sciences, Block-JD, Sector-III, Salt Lake, Kolkata 700106, India

*Email: sumanc@bose.res.in

**Abstract:** Estimating rare event kinetics from molecular dynamics (MD) simulations is a non-trivial task despite the great advances in enhanced sampling methods. Weighted Ensemble (WE) simulation, a special class of enhanced sampling technique, offers a way to directly calculate kinetic rate constants from biased trajectories without the need to modify the underlying energy landscape using bias potentials. Conventional WE algorithms use different binning schemes to partition the collective variable (CV) space separating the two metastable states of interest. In this work, we have developed a new "binless" WE simulation algorithm to bypass the hurdles of optimizing binning procedures. Our proposed protocol (WeTICA) uses a low dimensional CV space to drive the WE simulation towards the specified target state. We have applied this new algorithm to recover the unfolding kinetics of three proteins: A) TC5b Trp-cage mutant, B) TC10b Trp-cage mutant and C) Protein G with unfolding times spanning the range between $3\,\mu s - 40\,\mu s$ using projections along predefined fixed Time-lagged Independent Component Analysis (TICA) eigenvectors as CVs. Calculated unfolding times converge to the reported values with good accuracy with more than one order of magnitude less cumulative WE simulation time than the unfolding time scales with or without *a priori* knowledge of the CVs that can capture unfolding. Our algorithm can be used with other linear CVs, not limited to TICA. Moreover, the new walker selection criteria for resampling employed in this algorithm can be used on more sophisticated nonlinear CV space for further improvements of binless WE methods.

# I. Introduction

Molecular dynamics (MD) simulation is an indispensable tool to understand different biomolecular processes and calculate associated thermodynamic and kinetic properties, for example, binding energy of a ligand to a target receptor[1,2], free energy landscape of proteins[3] and kinetic rate constants associated with ligand dissociation (and association) from (to) a target receptor[4–6]. Despite the remarkable advances in MD software and hardware that enable us to access millisecond time scales at the atomistic resolution, normal MD simulations usually struggle to overcome the barriers associated with different processes making such events rare and difficult to capture.

A plethora of methods, known as enhanced sampling techniques, have been developed as a solution for sampling rare events. Some of the popular methods are Metadynamics (MetaD)[7–12] and its different variants[13], adaptive biasing force (ABF)[14–16], Gaussian-accelerated molecular dynamics (GaMD)[17,18], Replica exchange methods[19–23], $\tau$ −Random acceleration molecular dynamics ($\tau$RAMD)[24,25] and various methods employing machine learning based techniques[26–29]. These enhanced sampling methods not only facilitate quick exploration of the rugged biomolecular free energy landscape, but also enable the calculation of rare event kinetics[30–35] and building kinetic models like Markov state model (MSM)[36–39].

Weighted Ensemble (WE)[40–44] simulation is a special class of enhanced sampling method capable of calculating the kinetics of rare events. WE has proven to be useful in studying many biologically relevant systems[45–51]. WESTPA[52,53] is a popular toolkit to perform and analyse WE simulations. Generally, in conventional WE protocol, the configurational space is mapped onto a low dimensional collective variable (CV) space that describes the transition of interest, followed by dividing it into bins. Several trajectories (called "walkers" in WE framework) pre-assigned with probabilities or weights are simultaneously initiated from the initial structure. If $N_w$ trajectories are initiated, each walker will be assigned an initial weight $\frac{1}{N_w}$. Walkers that reach new bins are cloned, that is, two new simulations will be started from that configuration, and walkers residing in the same bin will be merged[54]. In this way, the simulations evolve under the natural dynamics of the system with trajectory resampling (cloning + merging) for many iterations to enhance the sampling of rare events. WE simulation protocol has been successfully combined with other methods like milestoning (WEM)[55,56], gaussian accelerated molecular dynamics (GaMD-WE)[57], neural networks (DeepWEST)[58] to improve thermodynamic and kinetic sampling. Since the resampling decisions are made based on the exploration of new bins in the CV space, the choice of appropriate CV space and binning schemes affect the algorithm's performance. Different binning techniques like voronoi polyhedral[59,60], finite temperature string method[61], minimal adaptive binning (MAB)[62], mean first passage time (MFPT) binning[63] have been developed to partition the CV space. Despite these mathematical developments, determination of good CV space and optimizing the binning scheme are still non-trivial tasks.

Resampling Ensembles by Variation Optimization or REVO[64,65] is a popular "binless" WE simulation algorithm based on the optimization of a quantity called "trajectory variation", which can be considered as a metric of how different the walkers are from each other. The use of binless WE method can bypass the hurdles of optimizing binning schemes to obtain quantitatively accurate and converged results. In a recent study[66], REVO was used to calculate residence times (in minutes time scales) of various inhibitors of soluble epoxide hydrolase (sEH). Despite the success of this algorithm, the current implementation of REVO in WE software Wepy[67] cannot be directly used to study other biologically relevant diverse class of problems like protein folding-unfolding or transition between different metastable conformational states of biomolecules.

To broaden the applicability of binless WE methods, in this work we have developed a new binless WE algorithm utilizing the fundamental principles of the REVO algorithm. Our algorithm (named WeTICA) has been implemented based on the Wepy[67] codebase. Our proposed protocol uses a fixed predefined low dimensional linear CV space to drive the WE simulation towards the specified target state. In this work, we have demonstrated the performance of this new algorithm using projections along Time-lagged Independent Component Analysis (TICA)[68] eigenvectors as CVs to recover the unfolding kinetics of three well studied proteins: 1) TC5b Trp-cage mutant, 2) TC10b Trp-cage mutant and 3) Protein G with known unfolding times spanning the range between 3 µs − 40 µs. Protein unfolding is a slow process and the choice of these three systems gives us a kinetic time scale window of reasonable size to test the performance of our new algorithm.

## II. Method

We have defined a new "trajectory variation" (V) function as follows:

$$V = \sum_i V_i = \sum_i \frac{1}{d_{from\ target}^i} \qquad (1)$$

where $d_{from\ target}^i$ is the distance of walker *i* measured from the target state conformation using some distance metric and the summation is over all the walkers at each cycle. Although the original algorithm of REVO used a different form for the variation function[64], our algorithm works similarly by maximizing this parameter *V* (Eq. 1) using trajectory resampling.

Root mean squared distance (RMSD) between protein backbone atoms is a very high dimensional distance metric routinely used to quantify the difference between two conformational states. However, the RMSD CV suffers from a significant degeneracy problem at larger values. Thus, having a large and same RMSD value of two walkers with respect to some distant reference state does not necessarily mean that they have the same conformation/state. This degeneracy problem becomes increasingly pronounced with higher dimensionality of the problem.

The choice of a lower dimensional CV space not only tackles the degeneracy issue associated with high dimensions, but choosing appropriate CVs can separate the metastable states of interest and capture the transition processes between the states. However, we should always keep in mind that any projection from a higher dimensional space to lower dimensional space has other demerits as well. Time-lagged Independent Component Analysis (TICA)[69,70] is capable of capturing slow modes in a molecular process. TICA is a linear dimensionality reduction method where the actual high dimensional descriptor data sets are projected on some leading TICA eigenvectors to visualize important metastable states in a low dimensional representation. Furthermore, TICA components are not only widely used to build MSMs to study kinetics from MD simulation data[71,72] , their use as CVs for enhanced sampling is also established[73]. Thus, we decided to project all the walkers after every cycle on some predefined TICA eigenvectors using some descriptor datasets and then calculate the Euclidean distances between the projection of the walkers and the projection of the target state conformation. So, the TICA eigenvectors serve as CVs in our algorithm. Note that the eigenvectors will be fixed throughout the entire simulation and due to the flexibility of this algorithm, any larger number of CVs (TICA eigenvectors) can be used without significant computational cost, unlike other popular enhanced sampling methods like umbrella sampling or metadynamics. Moreover, this algorithm can be used with any linear dimensionality reduction methods and is not just limited to TICA.

The $d^i_{\text{from target}}$ parameter in Eq. 1 is the Euclidean distance between the projection of the *i*-th walker and the projection of the target state conformation on this fixed TICA projection plane. We have implemented two different sets of input featurization schemes: 1) pair-wise distances between the $C_\alpha$ atoms, and 2) distances between any set of selected atom pairs. Apart from calculating $d^i_{\text{from target}}$, we also calculate distance between every pair of walkers $d_{ij}$ to make the merging decision based on some distance cutoff on the same TICA projection plane.

Since variation $V_i$ of the walker i is defined as the inverse of its distance from target (Eq.1), the walker that is closest to the target has the largest variation value. As a result, this algorithm selects the walker *i* which is closest to the target for cloning and the walker *j* which is farthest from the target as the first candidate for merging to fulfill the objective of maximizing total variation *V*. Then it will search for another walker *k* (excluding *i* and *j*) for the second candidate to form the merging pair with walker *j* if the distance between walker *j* and walker *k* is less than some predefined cut-off distance $d_{\text{merge}}$, such that $d_{jk} \leq d_{\text{merge}}$ and $w_j + w_k < p_{\text{max}}$. Here $w_j$ and $w_k$ are the weights of the walker *j* and walker *k*, respectively, and $p_{\text{max}}$ is the maximum weight that a walker can hold. We set $p_{\text{max}} = 0.20$ and any walker whose weight after cloning becomes less than $p_{\text{min}} = 10^{-12}$ is prohibited from cloning.

After selecting a suitable merging pair, a walker from that merging pair (*j*, *k*) is randomly selected for continuation with the total weight ($w_j + w_k$) and the other walker is discontinued. Like the original algorithm of REVO, this selection scheme goes on until V reaches a local maximum or no walkers are left for cloning and merging, at which point the ensemble is again propagated forward in time by the MD integrator.

To correctly calculate rate, it is necessary to recognise when the system has moved from the initial state basin and arrived in the desired one even if the precise knowledge of the corresponding transition state is not known[12]. We have used the target state conformation to steer the process to the correct direction. When a walker reaches inside a hypersphere of radius $d_{warp}$ centred around the projection of the target state conformation on the same projection plane, it is considered that the walker crosses into the final desired state basin. Then the corresponding weight of the walker will be saved and that walker will be re-initiated with the starting conformation. This process is called *warping* and the corresponding walker is called *warped* walker. Mean first passage time (MFPT) is calculated directly from the weights of the warped walkers as follows:

$$\text{MFPT} = \frac{T}{\sum_{(i,t) \in \mathcal{U}} w_{i,t}} \qquad (2)$$

where *T* is the total simulation time, $\mathcal{U}$ is a set of tuples denoting the walker indices *i* and the time point *t* when the walker is warped, $w_{i,t}$ is the weight of the warped walker *i* at that time point *t*. A full workflow diagram illustrating our algorithm is presented in Figure 1. Table S1 summarizes the key differences between the original REVO algorithm and our proposed algorithm.

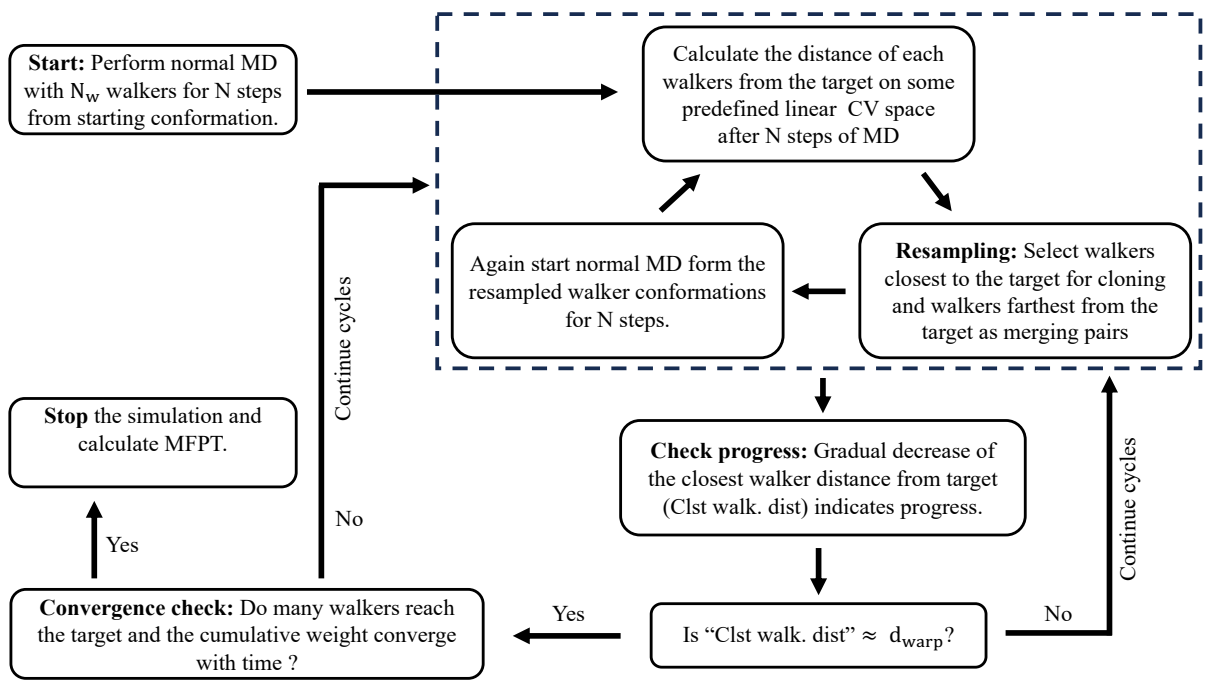

**Figure 1.** Workflow diagram illustrating the WeTICA algorithm.

## III. Guidelines for choosing WeTICA simulation parameters

Setting up a WeTICA simulation requires careful optimisation of certain input parameters to ensure efficient sampling of the productive trajectories. Here we provide some specific guidelines for choosing these parameters such as the number of walkers, generation and dimensionality of the CV space, walker merging distance $d_{merge}$ and the walker warping distance $d_{warp}$ that aim to an efficient and correct computation of MFPTs.

**A) Number of walkers:** Increasing the number of walkers will speed up the search process, but running greater number of simulations simultaneously will increase the computational wall time due to the limited number of computational resources. Thus, we advise to choose the number of walkers keeping a balance between these two factors.

**B) Generation and dimensionality of CV space:** The TICA eigenvectors should be computed before starting WeTICA simulations from unbiased MD trajectories (mostly in the two end states). As test cases, we have performed WeTICA simulations using TICA eigenvectors generated from both long unbiased trajectory and two short end-state simulation trajectories as discussed in the later sections of this paper. It turns out that this method is working in both cases.

Since the major goal of this kind of enhanced sampling methods is to compute the kinetics/rate of transition between the states, choose the minimum number of TICA eigenvectors as CVs that can ensure a proper separation of the two end states to avoid degeneracy arising due to large number of dimensions.

**C) The choice of $d_{merge}$:** Create a distribution using the distances between all the pair-wise projection points of the initial state ensemble conformations and then use the position of the minimum of this distance distribution as $d_{merge}$. This choice of $d_{merge}$ will resemble the approximate size of the initial state basin on that CV space and thus ensure that the merging walker pairs belong to the same basin.

**D) The choice of $d_{warp}$:** Create a distribution using the distances between all the pair-wise projection points of the target state ensemble conformations and then use the position of the minimum of this distance distribution as $d_{warp}$. This choice of $d_{warp}$ will resemble the approximate size of the target state basin on that CV space and thus ensure that the walkers have arrived in the target state basin when they are warped.

However, if the target state has high entropy (large number of possible configurations) with a flat basin in the free energy landscape, for example, targets in the protein unfolding or ligand unbinding processes, it might be difficult to sample enough data using short trajectories to obtain approximate size of the target state basin. In such cases, if the value for $d_{warp}$ is not carefully chosen, the walkers may diffuse randomly without entering the hypersphere of radius $d_{warp}$ centred around the representative target state conformation even though they have already reached the target state basin. This will slow down the convergence of the WeTICA simulation. In such cases, we advise to test various values of $d_{warp}$ while ensuring sufficient escape form the initial state basin and choose the largest value as $d_{warp}$ to speed up convergence and accurate calculation of rate constants.

In the upcoming sections, we will discuss about the simulation details for the various systems and the results obtained from the WeTICA simulations.

## IV. Simulation details

### A. TC5b mutant of Trp-cage

The NMR structure (PDB ID: 1L2Y[74]) of the 20-residue TC5b Trp-cage mutant was modelled using CHARMM36 force filed[75]. The protein was solvated in a cubic box of ~44 Å side length containing 2679 TIP3P[76] water molecules and one Cl⁻ ion to neutralize the system. The system was then energy minimized using the steepest descent algorithm and equilibrated for 200 ps in NPT ensemble at 300K temperature and 1 atm pressure using GROMACS v2019.6[77]. Production run was conducted in NPT ensemble for 1.85 μs at the same temperature and pressure. Temperature and pressure were maintained using velocity rescale method[78] with time constant of 0.1 ps and Parrinelo-Rahman barostat[79] with a time constant of 2 ps respectively. All simulations were performed under periodic boundary conditions and the long range electrostatic interactions were handled using the Particle Mesh Ewald (PME)[80] summation method. The cut-off distances for electrostatic and van der Waals interactions were set to 10 Å. Bonds containing hydrogen atoms were constrained with LINCS[81]. Leap-frog integrator was used with an integration time step of 2 fs. Frames were saved at every 20 ps interval. After the completion of the production simulation, fraction of native contacts (Q) were calculated using MDTraj python package[82] for all the frames of the entire 1.85 µs trajectory with respect to the first frame considered as the folded structure. A snapshot with Q ~ 0.2 was taken as a representative of the unfolded state ensemble. This representative unfolded conformation was used as the target state conformation for the subsequent WeTICA simulations. Folded and the representative unfolded structures for this Trp-cage mutant are shown in Figure 2a.

We choose 153 $C_\alpha - C_\alpha$ atom pair wise distances ($C_\alpha$ atoms are at least 2 residues apart i.e i and i+3) as the feature to calculate TICA eigenvectors trained on the entire 1.85 µs long folding-unfolding trajectory with a lagtime of 10 ns using PyEMMA v2.5.12 Python package[83]. The first two TICA eigenvectors (TIC1 and TIC2) were used as CVs to perform several independent WeTICA simulations in NPT ensemble starting from the solvated folded structure. Details of the WeTICA simulation parameters are provided in Table 1.

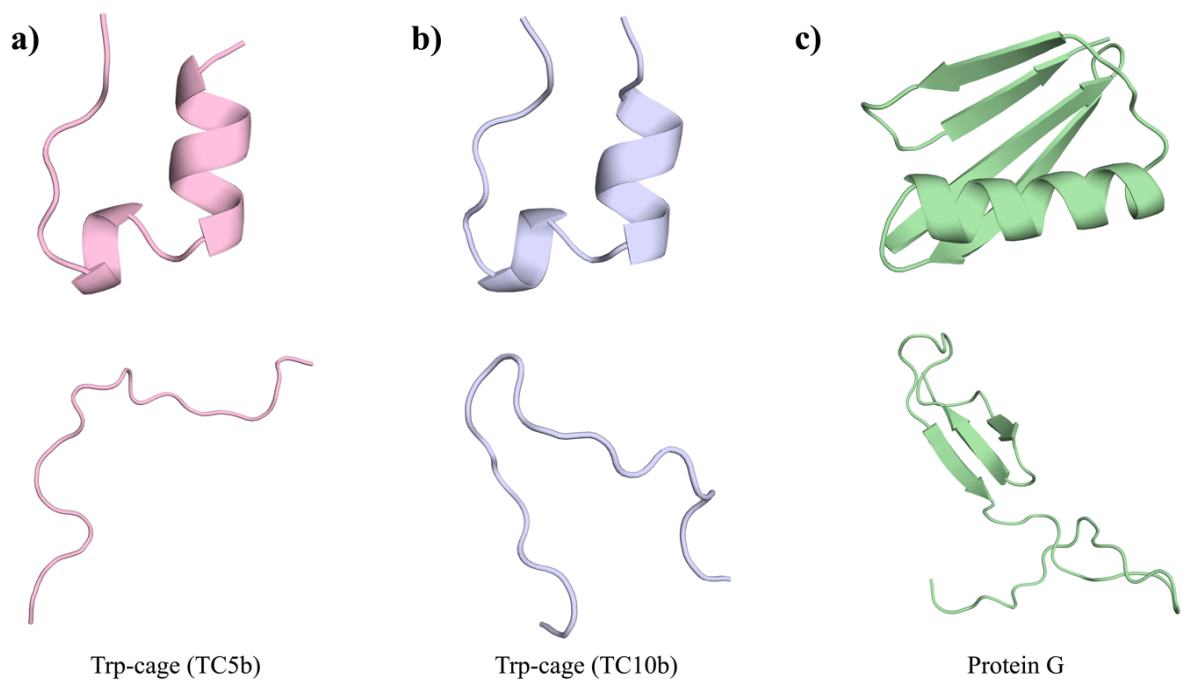

| | | |
|---|---|---|
| Trp-cage (TC5b) | Trp-cage (TC10b) | Protein G |

**Figure 2.** Upper row shows the folded structures and the lower row shows the representative unfolded structures in cartoon representation for **a)** TC5b Trp-cage mutant **b)** TC10b K8A Trp-cage mutant and **c)** Protein G, respectively. These unfolded structures were used as the target state conformations in the corresponding WeTICA simulations.

## B. TC10b mutant of Trp-cage

We took the 208 μs long explicit solvent K8A mutant of the 20-residue Trp-cage mini protein (TC10b) simulation trajectory generated by D.E Shaw Research[84]. The closest experimental structure is PDB ID: 2JOF[85]. The protein was modelled using CHARMM22* force field[86] and solvated with TIP3P[76] water molecules. Details about the simulation are provided in the supplementary material. We calculated the Q values of every frames of this trajectory with respect to the first frame considered as the folded structure. A snapshot with Q ~ 0.2 was taken as a representative of the unfolded state. This representative unfolded conformation was used as the target state conformation for the subsequent WeTICA simulations. Folded and the representative unfolded structures for this Trp-cage mutant are shown in Figure 2b.

This folded structure was solvated in a cubic box of ∼ 43 Å side length with 2505 TIP3P[76] water molecules, 0.65 mM NaCl concentration and modelled using CHARMM22* force field[86] with Asp & Arg side chains in their charged states. Again a set of 153 $C_\alpha - C_\alpha$ atom pair wise distances was used as the feature to calculate TICA eigenvectors trained on the entire 208 μs long Anton folding-unfolding trajectory of this mutant using a lagtime of 10 ns. First two TICA eigenvectors were used as CVs to perform several independent WeTICA simulations in NVT ensemble starting from the solvated folded structure. Details of the WeTICA simulation parameters for this Trp-cage mutant are provided in Table 1.

### C. Protein G

We took two trajectory segments of 2 μs long sampled the folded and unfolded states from the 168 μs long explicit solvent N37A/A46D/D77A triple mutant of the redesigned protein G variant NuG2 (closest experimental structure PDB ID:1MI0[87]) simulation trajectory generated by D.E Shaw Research[84]. The protein was modelled using CHARMM22* force field[86] and solvated with TIP3P[76] water molecules. Details about the simulation are provided in supplementary material. We calculated the Q values of every frames of these two trajectories. A snapshot with Q ~ 0.2 was taken as a representative of the unfolded state. This representative unfolded conformation was used as the target state conformation for the subsequent WeTICA simulations. Folded and the representative unfolded structures are shown in Figure 2c.

TICA eigenvectors were calculated using $C_\alpha - C_\alpha$ atom pair wise distances as the feature for a lagtime of 20 ns trained on these two end-state trajectories. The folded structure was modelled using CHARMM22* force field[86] with Asp, Glu and Lys side chains in their charged states and solvated with 100 mM NaCl salt concentration in a cubic box of ∼ 57 Å side length with 5438 TIP3P[76] water molecules. The solvated structure was energy minimized and equilibrated at 350K temperature. The first two eigenvectors were used to perform several independent WeTICA simulations in NVT ensemble from the equilibrated solvated folded structure. Details of the WeTICA simulation parameters are provided in Table 1.

All WE simulations were performed using a modified version of the Wepy[67] v1.1.0 software. Dynamics was performed using OpenMM v7.5.1[88]. Validation of the chosen WeTICA simulation parameters will be addressed in the next section while discussing results.

**Table 1**. Details of the WeTICA simulation parameters for the three systems.

| Parameters | Trp-cage (TC5b) | Trp-cage (TC10b) | Protein G |
|---|---|---|---|
| No. of walkers $N_w$ | 24 | 24 | 24 |
| Temperature | 300K | 290K | 350K |
| Feature | $C_\alpha - C_\alpha$ atom pair-wise distances | $C_\alpha - C_\alpha$ atom pair-wise distances | $C_\alpha - C_\alpha$ atom pair-wise distances |
| Eigenvectors | First two TICA eigenvectors. | First two TICA eigenvectors. | First two TICA eigenvectors. |
| Non-bonded cut-off | 10 Å | 9.0 Å | 9.5 Å |
| Integration time step | 2 fs | 2 fs | 2 fs |
| Resampling interval | 20 ps | 20 ps | 20 ps |
| Merge dist. $d_{merge}$ | 0.50 | 0.50 | 0.25 |
| Warp dist. $d_{warp}$ | 0.75 | 0.75 | 1.40 |

# V. Results & Discussions

## A. Unfolding kinetics of TC10b Trp-cage mutant

We first calculated the unfolding time ($\tau_u$) of the TC10b Trp-cage mutant using our protocol. Simulated unfolding time of $3 \pm 1$ μs at 290K temperature was reported from the direct analysis of the 208 μs long unbiased Anton trajectory[84] of this mutant as well as from a high resolution MSM built from the same trajectory[89]. Projections of all the frames of the entire Anton trajectory on the first two TICA eigenvectors (TIC1 and TIC2) used in WeTICA simulations were labelled with the $Q$ values of the corresponding conformations (see Figure S1). A clear separation between the folded and unfolded state ensembles on this projection plane validates the choice of this projection space to run the WeTICA simulations.

We set $d_{merge} = 0.50$ and $d_{warp} = 0.75$ for walker merging and warping, respectively. Notice that $d_{merge} = 0.50$ and $d_{warp} = 0.75$ correspond to the locations close to the first minima of the distance distributions between pairs of projection points for the folded and unfolded state basins, respectively (Figure S2). These cut-off distances were chosen to resemble the approximate size of the folded and unfolded state regions on this TICA space. Thus, this choice of the $d_{warp}$ ensures the arrival of the walkers in the unfolded state basin. In Figure 3a, we have shown the projections of a normal MD trajectory segment and WeTICA unfolding trajectory on the free energy surface (FES) derived using the projection of the full Anton trajectory. We can see that in normal MD the system got stuck in the folded state basin for 500 ns whereas in WeTICA simulation it has reached the unfolded region within ~ 6 ns. This highlights the sampling efficiency of our algorithm. In Figure 3b, we have plotted the $Q$ values of this WeTICA unfolding trajectory with time. In the inset of the Figure 3b, we have shown the distribution of the $Q$ values of the warped walker conformations. This $Q$ value distribution has a broad range (0.7 – 0.2) and also shows that all the walkers indeed left the folded state basin and crossed into the unfolded state region when they were warped by the algorithm to calculate unfolding kinetics. Note that due to the broad spread of the unfolded state basin as compared to the folded one (Figure 3a), larger values for $d_{warp}$ can be tested while ensuring proper escape of the walkers form the folded state basin ($Q \gtrsim 0.80$) to speed up convergence as discussed in Section III D. The same folded conformation was used in the fraction of native contact analysis for both normal MD and WeTICA unfolding trajectories.

The cumulative sum of the unfolding probabilities from each independent WeTICA simulations as well as the average result as a function of the total simulation time are plotted in Figure 3c. Average aggregated unfolding probability was used to subsequently calculate the unfolding time ($\tau_u$) of this Trp-cage mutant using Eq. 2. The final result is shown in Figure 3d. Unfolding time calculation converged to the reported value[84,89] ($3 \pm 1$ μs) within few "ns" of total simulation time. Computed unfolding time of $3.77 \pm 0.15$ μs (average and standard error of the data shown in the inset of Figure 3d) matches reasonably well with the previously reported simulated unfolding time. Therefore, our methodology not only enhances the sampling rate compared to unbiased MD, it also successfully reproduces the unfolding time with significantly less computational cost.

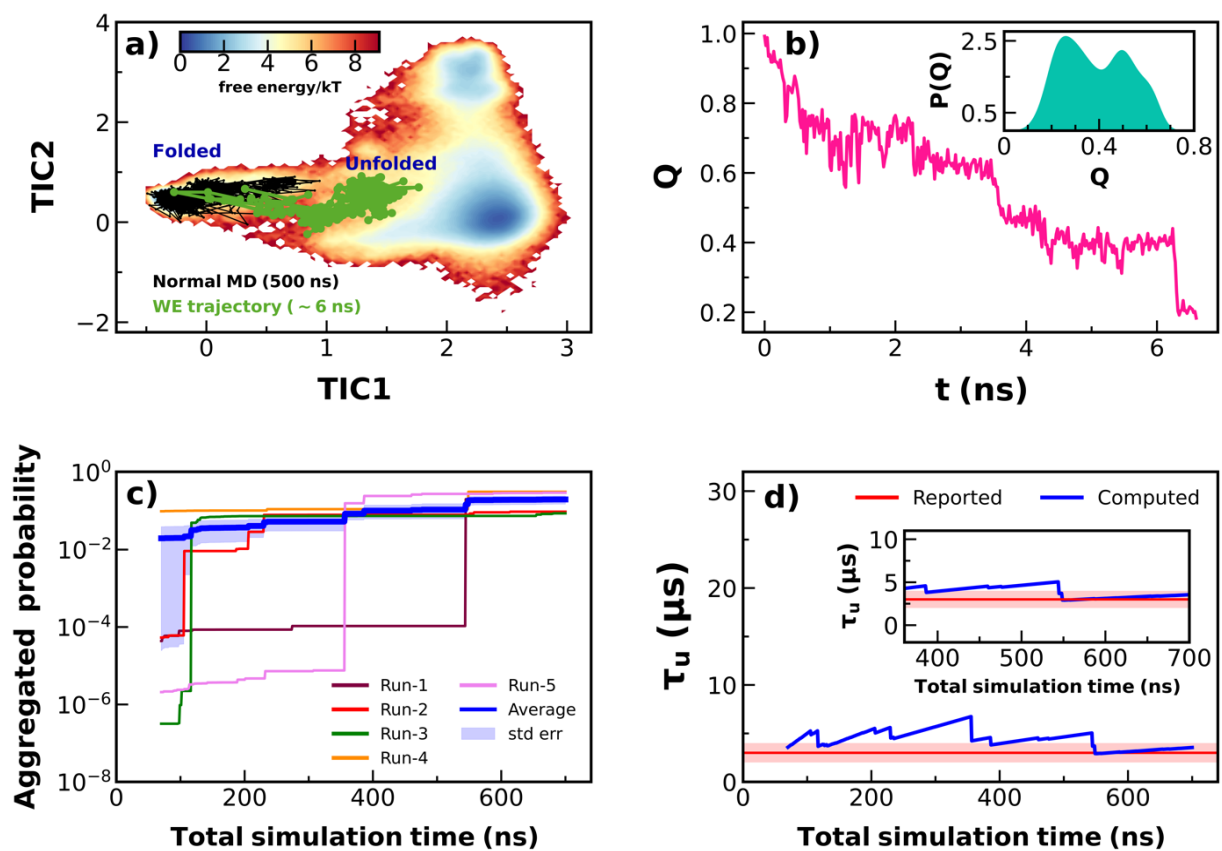

**Figure 3. a)** Projections of the normal MD trajectory segment and WeTICA unfolding trajectory on the first two TICA eigenvectors (TIC1 and TIC2) are shown on the free energy surface derived from the Anton trajectory. **b)** Fraction of native contacts ($Q$) with time are plotted for this WeTICA unfolding trajectory. Inset shows the distribution of the $Q$ values of the warped walker conformations. **c)** The cumulative sum of the unfolding probabilities from each independent WeTICA simulations as well as the average result as a function of the total simulation time are plotted. Blue shaded region represents the standard error of the mean (std err). **d)** Computed unfolding time ($\tau_u$) is plotted as a function of the total simulation time. Reported value (red line) represents the previously simulated unfolding time ($3 \pm 1$ μs) for the TC10b mutant of Trp-cage protein. Inset shows a close view of the converged region. Red shaded regions represent the standard error of the reported value.

## B. Unfolding kinetics of TC5b Trp-cage mutant

We next calculated the unfolding time ($\tau_u$) of the Trp-cage TC5b variant using our methodology. The experimental unfolding time of this mutant is 12.7 μs at 296K temperature[90]. Projections of all the frames of the 1.85 μs long normal MD trajectory on the first two TICA eigenvectors (TIC1 and TIC2) used in WeTICA simulations were labelled with the $Q$ values of the corresponding conformations (Figure S3). A clear separation between the folded and unfolded state basins for this system validates the use of this projection space for the WeTICA simulations.

We choose $d_{merge} = 0.50$ and $d_{warp} = 0.75$ for walker merging and warping, respectively. Note that $d_{merge} = 0.50$ and $d_{warp} = 0.75$ again correspond to the locations close to the positions of the first minima of the distance distributions between pairs of projection points for the folded and unfolded state basins, respectively (Figure S4). In Figure 4a, we have shown the comparison between a segment of normal MD trajectory and WeTICA unfolding trajectory on the free energy surface (FES) derived using the full normal MD trajectory. We can see that in normal MD the system got stuck in the folded basin for 500 ns whereas in WeTICA simulation the protein unfolds within ~ 20 ns. In Figure 4b, we have plotted the Q values of this WeTICA unfolding trajectory with time. In the inset of the Figure 4b, we have shown the distribution of the $Q$ values of the warped walker conformations. This distribution of the $Q$ values has a broad range (0.7 – 0.2) and also shows that all the walkers indeed left the folded state basin ($Q \gtrsim 0.80$) and arrived at the unfolded state basin when they were warped by the algorithm. As discussed in section III D, larger values for $d_{warp}$ can be chosen in this case also.

The cumulative sum of the unfolding probabilities from each independent WeTICA simulations as well as the average result as a function of the total simulation time are plotted in Figure 4c. Average aggregated unfolding probability was used to subsequently calculate the unfolding time ($\tau_u$) of this Trp-cage mutant. The final result is shown in Figure 4d. The calculated unfolding time converged to the experimental value[90] (12.7 μs) within at least one order of magnitude less cumulative WE simulation time than the unfolding time scale. Computed unfolding time of 12.87 ± 0.09 μs (average and standard error of the data shown in the inset of Figure 4d) matches reasonably well with the experimental unfolding time.

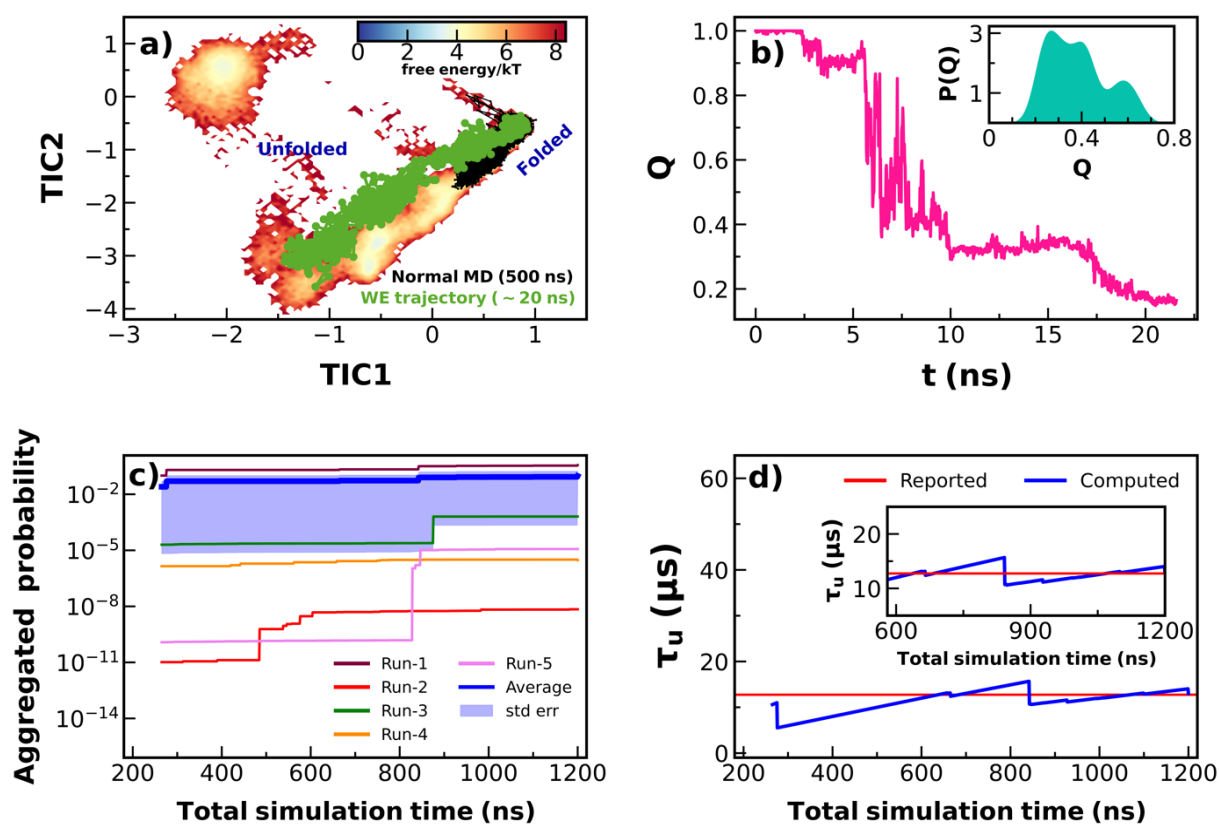

**Figure 4. a)** Projections of the normal MD trajectory segment and WeTICA unfolding trajectory on the first two TICA eigenvectors (TIC1 and TIC2) are shown on the free energy surface derived from the 1.85 μs long normal MD trajectory. **b)** Fraction of native contacts ($Q$) with time are plotted for this WeTICA unfolding trajectory. Inset shows the distribution of the *Q* values of the warped walker conformations. **c)** The cumulative sum of the unfolding probabilities from each independent WeTICA simulations as well as the average result as a function of the total simulation time are plotted. Blue shaded region represents standard error of the mean (std err). **d)** Computed unfolding time ($\tau_u$) is plotted as a function of the total simulation time. Reported value (red line) represents the experimental unfolding time (12.7 μs) for the TC5b mutant of Trp-cage protein. Inset shows a close view of the converged region.

For both the mutants of the Trp-cage mini protein, WeTICA simulations produced reasonably accurate unfolding times with only few "ns" of simulations. However, we calculated the TICA eigenvectors (CVs) for these two systems using long enough unbiased trajectories where several folding-unfolding events have been observed. Thus, by construction, the TICA eigenvectors are embedded with all the necessary information related to the slow folding-unfolding processes. But in most of the practical scenarios, the appropriate CVs that can capture the transition of interest are not known *a priori* due to the time scale limitation of unbiased MD simulation. On the other hand we can always generate TICA eigenvectors trained on two short end-state trajectories with perhaps no hopping between the two states. Now the question naturally arises: will our method be able to calculate rare event kinetics using CVs derived from such two end-state simulation trajectories? We will discuss this scenario using the example of the unfolding of Protein G in the next section.

## C. Unfolding kinetics of Protein G

Simulated unfolding time of 37 $\pm$ 10 μs at 350K temperature was reported from the direct analysis of several μs long Anton trajectories of redesigned Protein G[84] variant NuG2. The projections of the two Anton simulation trajectory segments belong to the folded and unfolded states on the first two TICA eigenvectors as used in WeTICA simulations are shown in Figure 5a. Each projection point is labelled according to the Q value of the corresponding conformation. Note that this time the folded and unfolded state regions are disconnected on this TICA projection plane as compared to the previous two examples. We set $d_{merge}$ = 0.25, which corresponds to the position of the minimum of the distance distribution between pairs of projection points for the folded state basin (Figure S6). As we have discussed in section III D and assuming that only 2 μs unbiased trajectory may not be sufficient to sample the whole unfolded state of large proteins like Protein G, we cannot confidently calculate the approximate size of the unfolded state region for large systems using short trajectories. Thus, we decided to check different values for the $d_{warp}$ parameter while ensuring proper escape of the walkers form the folded state basin (Q $\gtrsim$ 0.80). This is accomplished by monitoring the Q values of three productive WeTICA unfolding trajectories generated using three different values for $d_{warp}$ = 1.0, 1.2 and 1.4 with time. The results are shown in Figure 5b. We can see that in all three cases, the Q values of the walkers are within 0.7-0.6 when they were warped by the algorithm. This ensures that all the walkers have indeed left the folded state basin sufficiently before getting warped by the algorithm. Thus, we decided to take the maximum value, that is, $d_{warp}$ = 1.4 for all the independent WeTICA simulations of Protein G to speed up convergence and accurate calculation of unfolding MFPT.

The cumulative sum of the unfolding probabilities from each independent WeTICA simulations as well as the average result as a function of the total simulation time are plotted in Figure 5c. This average aggregated unfolding probability was used to subsequently calculate the unfolding time ($\tau_u$) of Protein G. The final result is shown in Figure 5d. The calculated unfolding time converged to the previously reported value[84] (37 $\pm$ 10 μs) within more than one order of magnitude less cumulative WE simulation time than the unfolding time scale.

Computed unfolding time of 47.35 ± 0.39 μs (average and standard error of the data shown in the inset of Figure 5d) also matches reasonably well with the reported value.

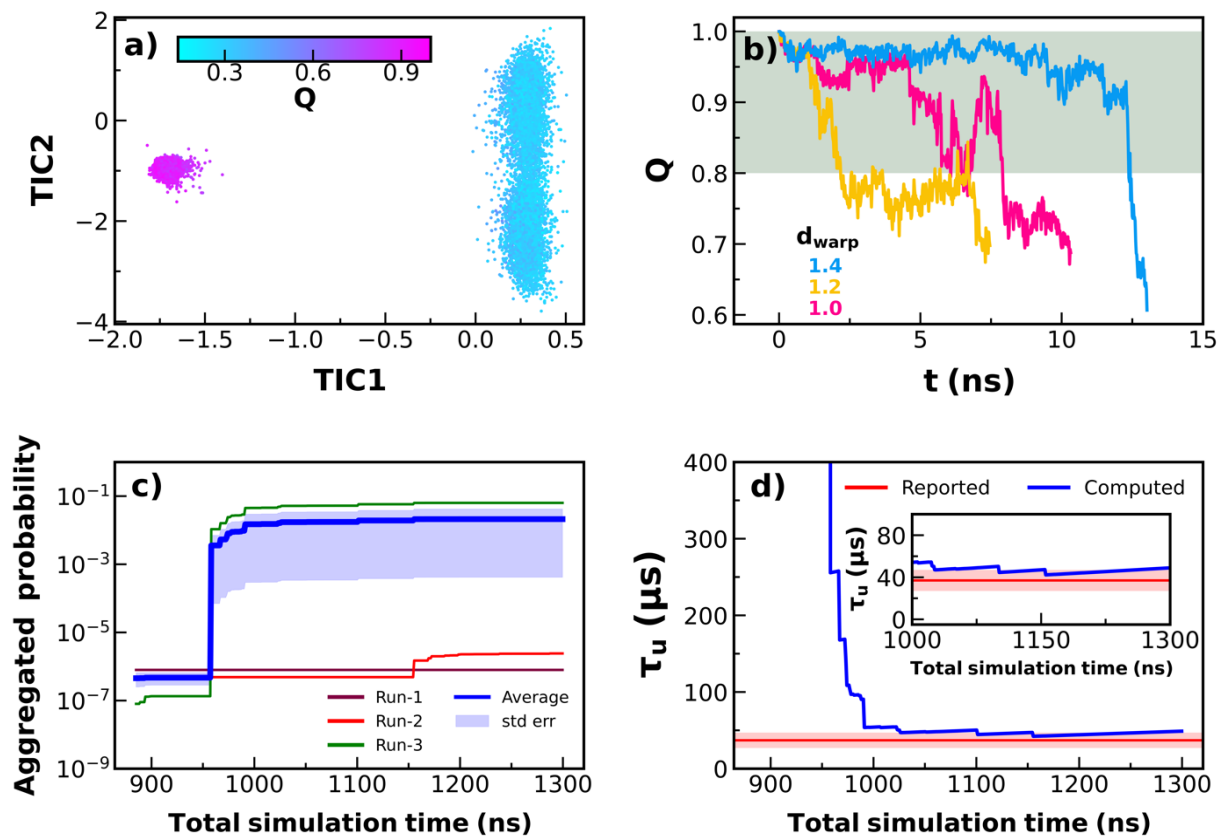

**Figure 5. a)** Frames from the two unbiased Anton trajectory segments (2 μs each) belong to the folded and unfolded states after projecting on the first two TICA eigenvectors (TIC1 and TIC2) are shown on a scatter plot. Each point is coloured according to the fraction of native contact (Q) value of the corresponding conformation. **b)** Fraction of native contacts (Q) of three productive WeTICA unfolding trajectories generated using three different values for $d_{warp}$ are plotted with time. The shaded region represents the folded basin (Q ≳ 0.80). **c)** The cumulative sum of the unfolding probabilities from each independent WeTICA simulations as well as the average result as a function of the total simulation time are plotted. Blue shaded region represents standard error of the mean (std err). **d)** Computed unfolding time ($\tau_u$) is plotted against the total simulation time. Reported value (red line) represents the previously simulated unfolding time (37 ± 10 µs). Inset shows a close view of the converged region. Red shaded regions represent the standard error (std err) of the reported value.

From all the above mentioned results for the three systems, we can conclude that our methodology significantly enhances the sampling and capable of computing kinetic rate constants with reasonable accuracy with or without a priori knowledge of correct CVs that can capture unfolding. The computed unfolding times for the three proteins are provided in Table-2 and several information, for example, total number of unfolding events and total number of WE cycles related to the independent WeTICA simulations for the three systems are provided in Table S2.

**Table-2.** Computed unfolding times ($\tau_u$) with standard errors for the three proteins are provided and compared with the previously reported simulated (Sim) and experimental (Expt) values.

| Systems | $\tau_u$ (µs) produced in this work | $\tau_u$ (µs) reported in literature[84,90] |
|---|---|---|
| TC10b Trp-cage mutant | 3.77 ± 0.15 | 3 ± 1 (Sim) |
| TC5b Trp-cage mutant | 12.87 ± 0.09 | 12.7 (Expt) |
| Protein G | 47.35 ± 0.39 | 37 ± 10 (Sim) |

# VI. Conclusion

In this work, we have developed a new "binless" WE simulation method named WeTICA utilizing the fundamental ideas from the REVO WE algorithm. Our proposed protocol uses fixed predefined linear CV space to drive the WE simulations towards the specified target state. We have demonstrated the performance of this new algorithm by recovering the unfolding kinetics of three proteins: 1) TC5b Trp-cage mutant, 2) TC10b Trp-cage mutant and 3) Protein G using Time-lagged Independent Component Analysis (TICA) eigenvectors as our predefined CVs.

For the two Trp-cage mutants, first we calculated TICA eigenvectors using long trajectories where several folding-unfolding events occurred. The first two TICA eigenvectors were used as CVs to guide the WeTICA simulations toward the target unfolded state. The calculated unfolding times converged to the reported unfolding times (in μs order) within feasible cumulative simulation time (within few "ns") for both the mutants.

In the third case of Protein G, we generated TICA eigenvectors trained on two end-state trajectories where the system was trapped inside the folded and unfolded state basins with no hopping between the two states. Our protocol again successfully reproduced the kinetics of Protein G unfolding using TICA eigenvectors computed from these two end state trajectories within practical simulation time.

Thus, our proposed protocol is working with or without *a priori* knowledge of the CVs that can capture the transitions of interest (in this case, unfolding). Although we used TICA to construct our CV space, generality of our algorithm allow the use of any linear projection methods e.g recently developed harmonic linear discriminant analysis (HLDA), which was proven as good CV for studying transitions between two metastable states[91,92]. The performance of our method can be tested with this kind of newly developed linear CVs in future. Moreover, this new way of walker selection for resampling can also be used on more sophisticated nonlinear CV space e.g variational autoencoder (VAE) latent space[93,94] for further improvements of binless WE methods. We believe that the generality and efficiency of the WE algorithm presented here will be helpful to study kinetics of a diverse class of biologically relevant problems.

## Supplementary material

The supplementary material contains major differences between the original REVO and our method, the simulation details of the Anton trajectories for TC10b Trp-cage mutant and protein G, and distance distributions between the projections of the normal MD trajectory for the folded and unfolded state basins on the TICA planes.

## Acknowledgments

This work has been partially funded by Science and Engineering Research Board (SERB), Government of India (project no. MTR/2021/000859). Authors acknowledge the supercomputing facility "Param Rudra" established under the National Supercomputing Mission (NSM), Government of India and the TRC computing facility in SNBNCBS, Kolkata. S.M. acknowledges SNBNCBS for providing research fellowship.

## Data availability

All the data required to reproduce our results and the python codes to run WeTICA are available at https://github.com/TeamSuman/WeTICA .

# References:

(1) Woo, H.-J.; Roux, B. Calculation of Absolute Protein--Ligand Binding Free Energy from Computer Simulations. Proceedings of the National Academy of Sciences 2005, 102 (19), 6825–6830.

(2) Duan, L.; Liu, X.; Zhang, J. Z. H. Interaction Entropy: A New Paradigm for Highly Efficient and Reliable Computation of Protein--Ligand Binding Free Energy. J Am Chem Soc 2016, 138 (17), 5722–5728.

(3) Iida, S.; Nakamura, H.; Higo, J. Enhanced Conformational Sampling to Visualize a Free-Energy Landscape of Protein Complex Formation. Biochemical Journal 2016, 473 (12), 1651–1662.

(4) Paul, F.; Wehmeyer, C.; Abualrous, E. T.; Wu, H.; Crabtree, M. D.; Schoneberg, J.; others. Protein-Peptide Association Kinetics beyond the Seconds Timescale from Atomistic Simulations. Nat Commun. 2017; 8 (1): 1095.

(5) Iida, S.; Kameda, T. Dissociation Rate Calculation via Constant-Force Steered Molecular Dynamics Simulation. J Chem Inf Model 2023, 63 (11), 3369–3376.

(6) Tiwary, P.; Limongelli, V.; Salvalaglio, M.; Parrinello, M. Kinetics of Protein-Ligand Unbinding: Predicting Pathways, Rates, and Rate-Limiting Steps. Proc Natl Acad Sci U S A 2015, 112 (5), E386–E391.

(7) Barducci, A.; Bonomi, M.; Parrinello, M. Metadynamics. Wiley Interdiscip Rev Comput Mol Sci 2011, 1 (5), 826–843.

(8) Laio, A.; Rodriguez-Fortea, A.; Gervasio, F. L.; Ceccarelli, M.; Parrinello, M. Assessing the Accuracy of Metadynamics. J Phys Chem B 2005, 109 (14), 6714–6721.

(9) Bussi, G.; Laio, A.; Parrinello, M. Equilibrium Free Energies from Nonequilibrium Metadynamics. Phys Rev Lett 2006, 96 (9), 90601.

(10) Barducci, A.; Bussi, G.; Parrinello, M. Well-Tempered Metadynamics: A Smoothly Converging and Tunable Free-Energy Method. Phys Rev Lett 2008, 100 (2), 20603.

(11) Laio, A.; Gervasio, F. L. Metadynamics: A Method to Simulate Rare Events and Reconstruct the Free Energy in Biophysics, Chemistry and Material Science. Reports on Progress in Physics 2008, 71 (12), 126601.

(12) Tiwary, P.; Parrinello, M. From Metadynamics to Dynamics. Phys Rev Lett 2013, 111 (23), 230602.

(13) Invernizzi, M.; Parrinello, M. Rethinking Metadynamics: From Bias Potentials to Probability Distributions. Journal of Physical Chemistry Letters 2020, 11 (7), 2731–2736.

(14) Darve, E.; Rodriguez-Gómez, D.; Pohorille, A. Adaptive Biasing Force Method for Scalar and Vector Free Energy Calculations. J Chem Phys 2008, 128 (14).

(15) Comer, J.; Gumbart, J. C.; Hénin, J.; Lelièvre, T.; Pohorille, A.; Chipot, C. The Adaptive Biasing Force Method: Everything You Always Wanted to Know but Were Afraid to Ask. J Phys Chem B 2015, 119 (3), 1129–1151.

(16) Darve, E.; Pohorille, A. Calculating Free Energies Using Average Force. J Chem Phys 2001, 115 (20), 9169–9183.

(17) Wang, J.; Arantes, P. R.; Bhattarai, A.; Hsu, R. V; Pawnikar, S.; Huang, Y. M.; Palermo, G.; Miao, Y. Gaussian Accelerated Molecular Dynamics: Principles and Applications. Wiley Interdiscip Rev Comput Mol Sci 2021, 11 (5), e1521.

(18) Miao, Y.; Feher, V. A.; McCammon, J. A. Gaussian Accelerated Molecular Dynamics: Unconstrained Enhanced Sampling and Free Energy Calculation. J Chem Theory Comput 2015, 11 (8), 3584–3595.

(19) Sugita, Y.; Okamoto, Y. Replica-Exchange Molecular Dynamics Method for Protein Folding. Chem Phys Lett 1999, 314 (1–2), 141–151.

(20) Zhang, W.; Wu, C.; Duan, Y. Convergence of Replica Exchange Molecular Dynamics. J Chem Phys 2005, 123 (15).

(21) Sindhikara, D.; Meng, Y.; Roitberg, A. E. Exchange Frequency in Replica Exchange Molecular Dynamics. J Chem Phys 2008, 128 (2).

(22) Rosta, E.; Hummer, G. Error and Efficiency of Replica Exchange Molecular Dynamics Simulations. J Chem Phys 2009, 131 (16).

(23) Wang, L.; Friesner, R. A.; Berne, B. J. Replica Exchange with Solute Scaling: A More Efficient Version of Replica Exchange with Solute Tempering (REST2). Journal of Physical Chemistry B 2011, 115 (30), 9431–9438.

(24) Kokh, D. B.; Amaral, M.; Bomke, J.; Grädler, U.; Musil, D.; Buchstaller, H.-P.; Dreyer, M. K.; Frech, M.; Lowinski, M.; Vallee, F.; others. Estimation of Drug-Target Residence Times by τ-Random Acceleration Molecular Dynamics Simulations. J Chem Theory Comput 2018, 14 (7), 3859–3869.

(25) Kokh, D. B.; Doser, B.; Richter, S.; Ormersbach, F.; Cheng, X.; Wade, R. C. A Workflow for Exploring Ligand Dissociation from a Macromolecule: Efficient Random

Acceleration Molecular Dynamics Simulation and Interaction Fingerprint Analysis of Ligand Trajectories. J Chem Phys 2020, 153 (12).

(26) Sultan, M. M.; Wayment-Steele, H. K.; Pande, V. S. Transferable Neural Networks for Enhanced Sampling of Protein Dynamics. J. Chem. Theory Comput. 2018, 14 (4), 1887–1894.

(27) Ribeiro, J. M. L.; Bravo, P.; Wang, Y.; Tiwary, P. Reweighted Autoencoded Variational Bayes for Enhanced Sampling (RAVE). Journal of Chemical Physics 2018, 149 (7), 72301.

(28) Chen, W.; Ferguson, A. L. Molecular Enhanced Sampling with Autoencoders: On-the-Fly Collective Variable Discovery and Accelerated Free Energy Landscape Exploration. J Comput Chem 2018, 39 (25), 2079–2102.

(29) Mehdi, S.; Smith, Z.; Herron, L.; Zou, Z.; Tiwary, P. Enhanced Sampling with Machine Learning. Annu Rev Phys Chem 2024, 75 (1), 347–370.

(30) Casasnovas, R.; Limongelli, V.; Tiwary, P.; Carloni, P.; Parrinello, M. Unbinding Kinetics of a P38 MAP Kinase Type II Inhibitor from Metadynamics Simulations. J Am Chem Soc 2017, 139 (13), 4780–4788.

(31) Wang, Y.; Ribeiro, J. M. L.; Tiwary, P. Past–Future Information Bottleneck for Sampling Molecular Reaction Coordinate Simultaneously with Thermodynamics and Kinetics. Nature Communications 2019 10:1 2019, 10 (1), 1–8.

(32) Best, R. B.; Hummer, G. Reaction Coordinates and Rates from Transition Paths. Proc. Natl. Acad. Sci. USA 2005, 102 (19), 6732–6737.

(33) Wang, Y.; Valsson, O.; Tiwary, P.; Parrinello, M.; Lindorff-Larsen, K. Frequency Adaptive Metadynamics for the Calculation of Rare-Event Kinetics. Journal of Chemical Physics 2018, 149 (7), 72309.

(34) Ray, D.; Ansari, N.; Rizzi, V.; Invernizzi, M.; Parrinello, M. Rare Event Kinetics from Adaptive Bias Enhanced Sampling. J Chem Theory Comput 2022, 18 (11), 6500–6509.

(35) Ray, D.; Parrinello, M. Kinetics from Metadynamics: Principles, Applications, and Outlook. J Chem Theory Comput 2023, 19 (17), 5649–5670.

(36) Prinz, J.-H.; Wu, H.; Sarich, M.; Keller, B.; Senne, M.; Held, M.; Chodera, J. D.; Schütte, C.; Noé, F. Markov Models of Molecular Kinetics: Generation and Validation. J Chem Phys 2011, 134 (17).

(37) Husic, B. E.; Pande, V. S. Markov State Models: From an Art to a Science. J Am Chem Soc 2018, 140 (7), 2386–2396.

(38) Pande, V. S.; Beauchamp, K.; Bowman, G. R. Everything You Wanted to Know about Markov State Models but Were Afraid to Ask. Methods 2010, 52 (1), 99–105.

(39) Prinz, J.-H.; Keller, B.; Noé, F. Probing Molecular Kinetics with Markov Models: Metastable States, Transition Pathways and Spectroscopic Observables. Physical Chemistry Chemical Physics 2011, 13 (38), 16912–16927.

(40) Huber, G. A.; Kim, S. Weighted-Ensemble Brownian Dynamics Simulations for Protein Association Reactions. Biophys J 1996, 70 (1), 97–110.

(41) Bhatt, D.; Zhang, B. W.; Zuckerman, D. M. Steady-State Simulations Using Weighted Ensemble Path Sampling. J Chem Phys 2010, 133 (1).

(42) Abdul-Wahid, B.; Feng, H.; Rajan, D.; Costaouec, R.; Darve, E.; Thain, D.; Izaguirre, J. A. AWE-WQ: Fast-Forwarding Molecular Dynamics Using the Accelerated Weighted Ensemble. J Chem Inf Model 2014, 54 (10), 3033–3043.

(43) Copperman, J.; Zuckerman, D. M. Accelerated Estimation of Long-Timescale Kinetics from Weighted Ensemble Simulation via Non-Markovian "Microbin" Analysis. J Chem Theory Comput 2020, 16 (11), 6763–6775.

(44) Hellemann, E.; Durrant, J. D. Worth the Weight: Sub-Pocket EXplorer (SubPEx), a Weighted Ensemble Method to Enhance Binding-Pocket Conformational Sampling. J Chem Theory Comput 2023, 19 (17), 5677–5689.

(45) Adelman, J. L.; Grabe, M. Simulating Current--Voltage Relationships for a Narrow Ion Channel Using the Weighted Ensemble Method. J Chem Theory Comput 2015, 11 (4), 1907–1918.

(46) Adhikari, U.; Mostofian, B.; Copperman, J.; Subramanian, S. R.; Petersen, A. A.; Zuckerman, D. M. Computational Estimation of Microsecond to Second Atomistic Folding Times. J Am Chem Soc 2019, 141 (16), 6519–6526.

(47) Saglam, A. S.; Chong, L. T. Protein--Protein Binding Pathways and Calculations of Rate Constants Using Fully-Continuous, Explicit-Solvent Simulations. Chem Sci 2019, 10 (8), 2360–2372.

(48) Lotz, S. D.; Dickson, A. Unbiased Molecular Dynamics of 11 Min Timescale Drug Unbinding Reveals Transition State Stabilizing Interactions. J Am Chem Soc 2018, 140 (2), 618–628.

(49) Dixon, T.; Lotz, S. D.; Dickson, A. Predicting Ligand Binding Affinity Using On-and off-Rates for the SAMPL6 SAMPLing Challenge. J Comput Aided Mol Des 2018, 32 (10), 1001–1012.

(50) Sztain, T.; Ahn, S.-H.; Bogetti, A. T.; Casalino, L.; Goldsmith, J. A.; Seitz, E.; McCool, R. S.; Kearns, F. L.; Acosta-Reyes, F.; Maji, S.; others. A Glycan Gate Controls Opening of the SARS-CoV-2 Spike Protein. Nat Chem 2021, 13 (10), 963–968.

(51) Ahn, S.-H.; Jagger, B. R.; Amaro, R. E. Ranking of Ligand Binding Kinetics Using a Weighted Ensemble Approach and Comparison with a Multiscale Milestoning Approach. J Chem Inf Model 2020, 60 (11), 5340–5352.

(52) Zwier, M. C.; Adelman, J. L.; Kaus, J. W.; Pratt, A. J.; Wong, K. F.; Rego, N. B.; Suárez, E.; Lettieri, S.; Wang, D. W.; Grabe, M.; others. WESTPA: An Interoperable, Highly Scalable Software Package for Weighted Ensemble Simulation and Analysis. J Chem Theory Comput 2015, 11 (2), 800–809.

(53) Russo, J. D.; Zhang, S.; Leung, J. M. G.; Bogetti, A. T.; Thompson, J. P.; Degrave, A. J.; Torrillo, P. A.; Pratt, A. J.; Wong, K. F.; Xia, J.; Copperman, J.; Adelman, J. L.; Zwier, M. C.; Lebard, D. N.; Zuckerman, D. M.; Chong, L. T. WESTPA 2.0: High-Performance Upgrades for Weighted Ensemble Simulations and Analysis of Longer-Timescale Applications. J Chem Theory Comput 2022, 18 (2), 638–649.

(54) Zuckerman, D. M.; Chong, L. T. Weighted Ensemble Simulation: Review of Methodology, Applications, and Software. Annu Rev Biophys 2017, 46, 43–57.

(55) Ray, D.; Andricioaei, I. Weighted Ensemble Milestoning (WEM): A Combined Approach for Rare Event Simulations. Journal of Chemical Physics 2020, 152 (23).

(56) Ray, D.; Stone, S. E.; Andricioaei, I. Markovian Weighted Ensemble Milestoning (M-WEM): Long-Time Kinetics from Short Trajectories. J Chem Theory Comput 2022, 18 (1), 79–95.

(57) Ahn, S.-H.; Ojha, A. A.; Amaro, R. E.; McCammon, J. A. Gaussian-Accelerated Molecular Dynamics with the Weighted Ensemble Method: A Hybrid Method Improves Thermodynamic and Kinetic Sampling. J Chem Theory Comput 2021, 17 (12), 7938–7951.

(58) Ojha, A. A.; Thakur, S.; Ahn, S. H.; Amaro, R. E. DeepWEST: Deep Learning of Kinetic Models with the Weighted Ensemble Simulation Toolkit for Enhanced Sampling. J Chem Theory Comput 2023, 19 (4), 1342–1359.

(59) Zhang, B. W.; Jasnow, D.; Zuckerman, D. M. The "Weighted Ensemble" Path Sampling Method Is Statistically Exact for a Broad Class of Stochastic Processes and Binning Procedures. J Chem Phys 2010, 132 (5).

(60) Dickson, A.; Brooks III, C. L. WExplore: Hierarchical Exploration of High-Dimensional Spaces Using the Weighted Ensemble Algorithm. J Phys Chem B 2014, 118 (13), 3532–3542.

(61) Adelman, J. L.; Grabe, M. Simulating Rare Events Using a Weighted Ensemble-Based String Method. Journal of Chemical Physics 2013, 138 (4), 44105.

(62) Torrillo, P. A.; Bogetti, A. T.; Chong, L. T. A Minimal, Adaptive Binning Scheme for Weighted Ensemble Simulations. Journal of Physical Chemistry A 2021, 125 (7), 1642–1649.

(63) Aristoff, D.; Copperman, J.; Simpson, G.; Webber, R. J.; Zuckerman, D. M. Weighted Ensemble: Recent Mathematical Developments. Journal of Chemical Physics 2023, 158 (1).

(64) Donyapour, N.; Roussey, N. M.; Dickson, A. REVO: Resampling of Ensembles by Variation Optimization. J Chem Phys 2019, 150 (24).

(65) Roussey, N. M.; Dickson, A. Quality over Quantity: Sampling High Probability Rare Events with the Weighted Ensemble Algorithm. J Comput Chem 2023, 44 (8), 935–947.

(66) Bose, S.; Lotz, S. D.; Deb, I.; Shuck, M.; Lee, K. S. S.; Dickson, A. How Robust Is the Ligand Binding Transition State? J Am Chem Soc 2023, 145 (46), 25318–25331.

(67) Lotz, S. D.; Dickson, A. Wepy: A Flexible Software Framework for Simulating Rare Events with Weighted Ensemble Resampling. ACS Omega 2020, 5 (49), 31608–31623.

(68) Schultze, S.; Grubmüller, H. Time-Lagged Independent Component Analysis of Random Walks and Protein Dynamics. J Chem Theory Comput 2021, 17 (9), 5766–5776.

(69) Naritomi, Y.; Fuchigami, S. Slow Dynamics in Protein Fluctuations Revealed by Time-Structure Based Independent Component Analysis: The Case of Domain Motions. J Chem Phys 2011, 134 (6).

(70) Pérez-Hernández, G.; Noé, F. Hierarchical Time-Lagged Independent Component Analysis: Computing Slow Modes and Reaction Coordinates for Large Molecular Systems. J Chem Theory Comput 2016, 12 (12), 6118–6129.

(71) Pérez-Hernández, G.; Paul, F.; Giorgino, T.; De Fabritiis, G.; Noé, F. Identification of Slow Molecular Order Parameters for Markov Model Construction. J Chem Phys 2013, 139 (1).

(72) Schwantes, C. R.; Shukla, D.; Pande, V. S. Markov State Models and TICA Reveal a Nonnative Folding Nucleus in Simulations of NuG2. Biophys J 2016, 110 (8), 1716–1719.

(73) Sultan, M. M.; Pande, V. S. TICA-Metadynamics: Accelerating Metadynamics by Using Kinetically Selected Collective Variables. J Chem Theory Comput 2017, 13 (6), 2440–2447.

(74) Neidigh, J. W.; Fesinmeyer, R. M.; Andersen, N. H. Designing a 20-Residue Protein. Nature Structural Biology 2002 9:6 2002, 9 (6), 425–430.

(75) Huang, J.; Mackerell, A. D. CHARMM36 All-Atom Additive Protein Force Field: Validation Based on Comparison to NMR Data. J Comput Chem 2013, 34 (25), 2135–2145.

(76) Mark, P.; Nilsson, L. Structure and Dynamics of the TIP3P, SPC, and SPC/E Water Models at 298 K. Journal of Physical Chemistry A 2001, 105 (43), 9954–9960.

(77) Abraham, M. J.; Murtola, T.; Schulz, R.; Páll, S.; Smith, J. C.; Hess, B.; Lindahl, E. GROMACS: High Performance Molecular Simulations through Multi-Level Parallelism from Laptops to Supercomputers. SoftwareX 2015, 1-2, 19–25.

(78) Bussi, G.; Donadio, D.; Parrinello, M. Canonical Sampling through Velocity Rescaling. J Chem Phys 2007, 126 (1), 14101.

(79) Parrinello, M.; Rahman, A. Polymorphic Transitions in Single Crystals: A New Molecular Dynamics Method. J Appl Phys 1981, 52 (12), 7182–7190.

(80) Essmann, U.; Perera, L.; Berkowitz, M. L.; Darden, T.; Lee, H.; Pedersen, L. G. A Smooth Particle Mesh Ewald Method. J Chem Phys 1995, 103 (19), 8577–8593.

(81) Hess, B.; Bekker, H.; Berendsen, H. J. C.; Fraaije, J. G. E. M. LINCS: A Linear Constraint Solver for Molecular Simulations. J Comput Chem 1997, 18 (12), 1463–1472.

(82) McGibbon, R. T.; Beauchamp, K. A.; Harrigan, M. P.; Klein, C.; Swails, J. M.; Hernández, C. X.; Schwantes, C. R.; Wang, L.-P.; Lane, T. J.; Pande, V. S. MDTraj: A Modern Open Library for the Analysis of Molecular Dynamics Trajectories. Biophys J 2015, 109 (8), 1528–1532.

(83) Scherer, M. K.; Trendelkamp-Schroer, B.; Paul, F.; Pérez-Hernández, G.; Hoffmann, M.; Plattner, N.; Wehmeyer, C.; Prinz, J.-H.; Noé, F. PyEMMA 2: A Software Package for Estimation, Validation, and Analysis of Markov Models. J Chem Theory Comput 2015, 11 (11), 5525–5542.

(84) Lindorff-Larsen, K.; Piana, S.; Dror, R. O.; Shaw, D. E. How Fast-Folding Proteins Fold. Science (1979) 2011, 334 (6055), 517–520.

(85) Barua, B.; Lin, J. C.; Williams, V. D.; Kummler, P.; Neidigh, J. W.; Andersen, N. H. The Trp-Cage: Optimizing the Stability of a Globular Miniprotein. Protein Engineering, Design & Selection 2008, 21 (3), 171–185.

(86) Piana, S.; Lindorff-Larsen, K.; Shaw, D. E. How Robust Are Protein Folding Simulations with Respect to Force Field Parameterization? Biophys J 2011, 100 (9).

(87) Nauli, S.; Kuhlman, B.; Trong, I. Le; Stenkamp, R. E.; Teller, D.; Baker, D. Crystal Structures and Increased Stabilization of the Protein G Variants with Switched Folding Pathways NuG1 and NuG2. Protein Science 2002, 11 (12), 2924–2931.

(88) Eastman, P.; Swails, J.; Chodera, J. D.; McGibbon, R. T.; Zhao, Y.; Beauchamp, K. A.; Wang, L. P.; Simmonett, A. C.; Harrigan, M. P.; Stern, C. D.; Wiewiora, R. P.; Brooks, B. R.; Pande, V. S. OpenMM 7: Rapid Development of High Performance Algorithms for Molecular Dynamics. PLoS Comput Biol 2017, 13 (7), e1005659.

(89) Sidky, H.; Chen, W.; Ferguson, A. L. High-Resolution Markov State Models for the Dynamics of Trp-Cage Miniprotein Constructed over Slow Folding Modes Identified by State-Free Reversible VAMPnets. J Phys Chem B 2019, 123 (38), 7999–8009.

(90) Qiu, L.; Pabit, S. A.; Roitberg, A. E.; Hagen, S. J. Smaller and Faster: The 20-Residue Trp-Cage Protein Folds in 4 μs. J Am Chem Soc 2002, 124 (44), 12952–12953.

(91) Mendels, D.; Piccini, G.; Parrinello, M. Collective Variables from Local Fluctuations. Journal of Physical Chemistry Letters 2018, 9 (11), 2776–2781.

(92) Mendels, D.; Piccini, G.; Brotzakis, Z. F.; Yang, Y. I.; Parrinello, M. Folding a Small Protein Using Harmonic Linear Discriminant Analysis. Journal of Chemical Physics 2018, 149 (19), 194113.

(93) Mansoor, S.; Baek, M.; Park, H.; Lee, G. R.; Baker, D. Protein Ensemble Generation Through Variational Autoencoder Latent Space Sampling. J Chem Theory Comput 2024, 20 (7), 2689–2695.

(94) Belkacemi, Z.; Bianciotto, M.; Minoux, H.; Lelièvre, T.; Stoltz, G.; Gkeka, P. Autoencoders for Dimensionality Reduction in Molecular Dynamics: Collective Variable Dimension, Biasing, and Transition States. Journal of Chemical Physics 2023, 159 (2), 24122.

# Supplementary Material

## WeTICA: A Directed Search Weighted Ensemble Based Enhanced Sampling Method to Estimate Rare Event Kinetics in a Reduced Dimensional Space

Sudipta Mitra, Ranjit Biswas and Suman Chakrabarty*

Department of Chemical and Biological Sciences, S. N. Bose National Centre for Basic Sciences, Block-JD, Sector-III, Salt Lake, Kolkata 700106, India

*Email: sumanc@bose.res.in

**Table S1:** Key comparisons between the original REVO and our algorithm

| Original REVO algorithm | Our Algorithm |
|---|---|
| **1)** The current implementation is only limited to the study of ligand binding/unbinding and cannot be directly used for other diverse classes of problems like protein folding-unfolding or transition between different metastable states of biomolecules. | **1)** The goal is to generalize the scope of REVO-based binless methods to a diverse class of problems, not limited to ligand binding/unbinding. |
| **2)** Cloning and merging of walkers are decided based on ligand RMSD between a pair of walkers $d_{ij}$. | **2)** Cloning and merging are decided based on projections of the walkers on a lower dimensional CV space, where we calculate the Euclidean distance $d_{ij}$ between those projections of walkers and also their distance from the target conformation $d^i_{\text{from target}}$. |
| **3)** Variation $V$ is defined as $$V = \sum_i V_i = \sum_i \sum_j \left(\frac{d_{ij}}{d_0}\right)^\alpha \phi_i \phi_j$$ | **3)** Variation $V$ is defined as $$V = \sum_i V_i = \sum_i \frac{1}{d^i_{\text{from target}}}$$ |
| **4)** Clone that walker which is farthest from all other walkers. | **4)** Clone that walker which is closest to the target state conformation. |

**Table S2.** Information related to all the independent WeTICA simulations.

| System | Run | Number of warping events | Total number of cycles | Total simulation time (ns) |
|---|---|---|---|---|
| Trp-cage (TC10b) | 1 | 311 | 1461 | 701.28 |
| | 2 | 264 | 1460 | 700.80 |
| | 3 | 113 | 1465 | 703.20 |
| | 4 | 159 | 1466 | 703.40 |
| | 5 | 127 | 1470 | 705.60 |
| Trp-cage (TC5b) | 1 | 36 | 2502 | 1200.96 |
| | 2 | 48 | 2505 | 1202.40 |
| | 3 | 87 | 2515 | 1207.20 |
| | 4 | 48 | 2501 | 1200.48 |
| | 5 | 53 | 2498 | 1199.04 |
| Protein G | 1 | 113 | 2800 | 1344.00 |
| | 2 | 90 | 2805 | 1346.40 |
| | 3 | 126 | 2810 | 1348.80 |

# Simulation details of the TC10b Trp-cage and Protein G anton trajectories[1]

**TC10b Trp-cage mutant:**

The simulation we study is a 208 μs long explicit solvent K8A mutant of the 20-residue Trp-cage mini protein (TC10b) with sequence: ASP-ALA-TYR-ALA-GLN-TRP-LEU-ALA-ASP-GLY-GLY-PRO-SER-SER-GLY-ARG-PRO-PRO-PRO-SER performed by D.E Shaw Research. The protein was prepared with Asp & Arg side chains and N- and C- terminal in their charged states and modelled using CHARMM22* force field. The protein was solvated with 65mM NaCl concentration in a cubic box of ∼ 37 Å side length containing ∼ 1700 TIP3P water molecules. Production runs were conducted in NVT ensemble at 290K temperature using special purpose Anton hardware. Integration time step was 2.5 fs and temperature was maintained using Nose-Hoover thermostat with 1 ps time constant. Short-range electrostatic and Lennard-Jones interactions were treated using 9 Å cutoff and long-range electrostatic interactions were treated with the Gaussian Split Ewald (GSE) method and a 32 × 32 × 32 cubic grid. Frames were saved at every 200 ps interval.

**Protein G:**

The simulation we study is a 168 μs long explicit solvent N37A/A46D/D77A triple mutant of the redesigned protein G variant NuG2 with sequence: ASP-THR-TYR-LYS-LEU-VAL-ILE- VAL-LEU-ASN-GLY-THR-THR-PHE-THR-TYR-THR-THR-GLU-ALA-VAL-ASP-ALA-ALA-THR-ALA-GLU-LYS-VAL-PHE-LYS-GLN-TYR-ALA-ASN-ASP-ALA-GLY-VAL-ASP-GLY-GLU-TRP-THR-TYR-ASP-ALA-ALA-THR-LYS-THR-PHE-THR-VAL-THR- GLU performed by D.E Shaw Research. The protein was prepared with Asp, Glu and Lys side chains and N- and C- terminal in their charged states and modelled using CHARMM22* force field. The protein was solvated with 100 mM NaCl concentration in a cubic box of ∼ 55 Å side length containing ∼ 5100 TIP3P water molecules. Production runs were conducted in NVT ensemble at 350K temperature using special purpose Anton hardware. Integration time step was 2.5 fs and temperature was maintained using Nose-Hoover thermostat with 1 ps time constant. Short-range electrostatic and Lennard-Jones interactions were treated using 9.5 Å cutoff . Frames were saved at every 200 ps interval.

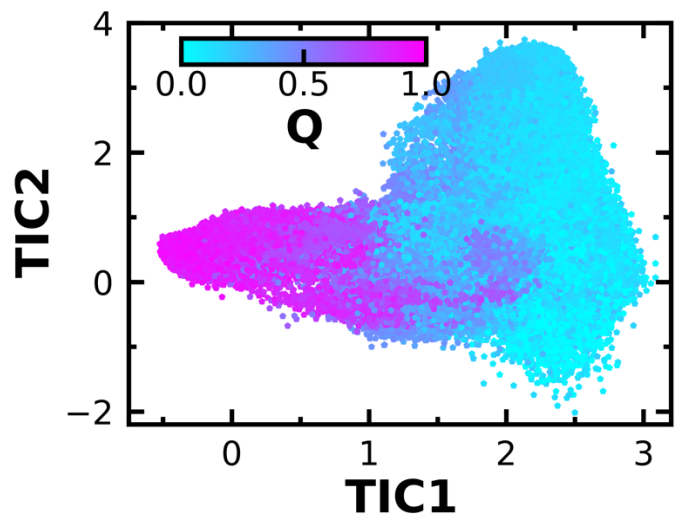

**Figure S1.** Frames of the Anton trajectory of the TC10b Trp-cage mutant after projecting on the first two TICA eigenvectors (TIC1 and TIC2) are shown as a scatter plot. Each point is coloured according to the fraction of native contact (Q) value of the corresponding conformation.

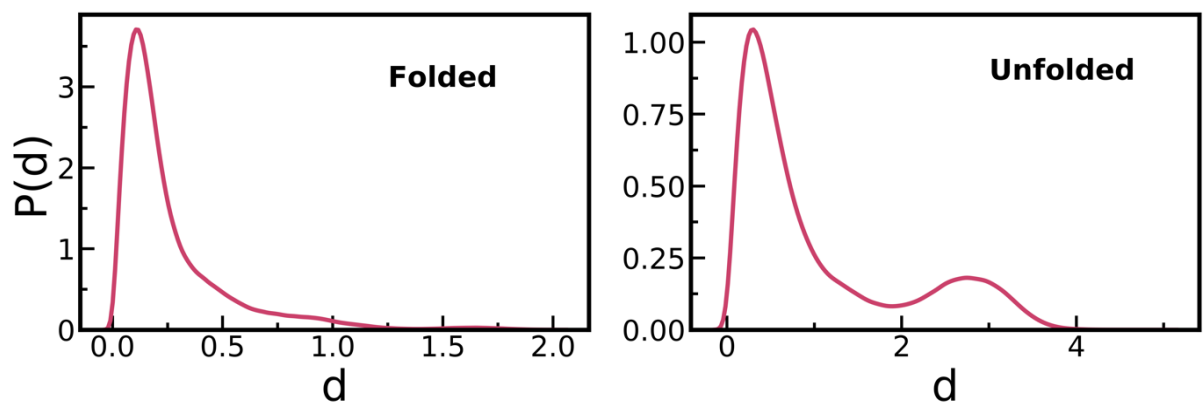

**Figure S2:** Distribution of pairwise distances between the projection points of the Anton trajectory frames on the TIC1 vs TIC2 plane in the folded and unfolded state regions for TC10b Trp-cage mutant.

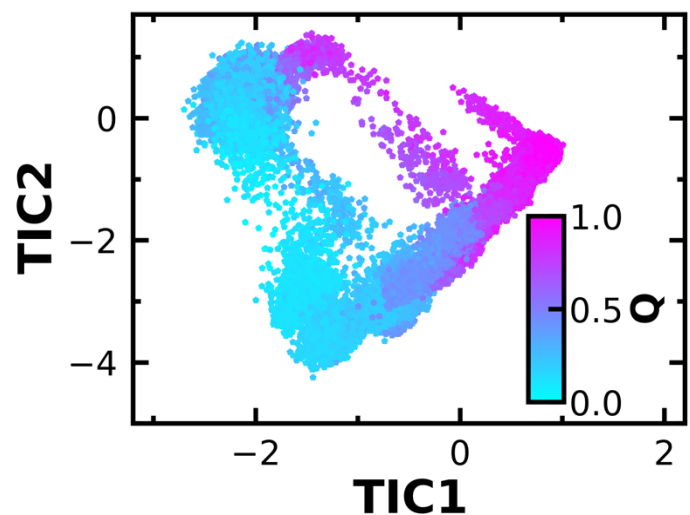

**Figure S3.** Frames of the normal MD trajectory of the TC5b Trp-cage mutant after projecting on the first two TICA eigenvectors (TIC1 and TIC2) are shown as a scatter plot. Each point is coloured according to the fraction of native contact (Q) value of the corresponding conformation.

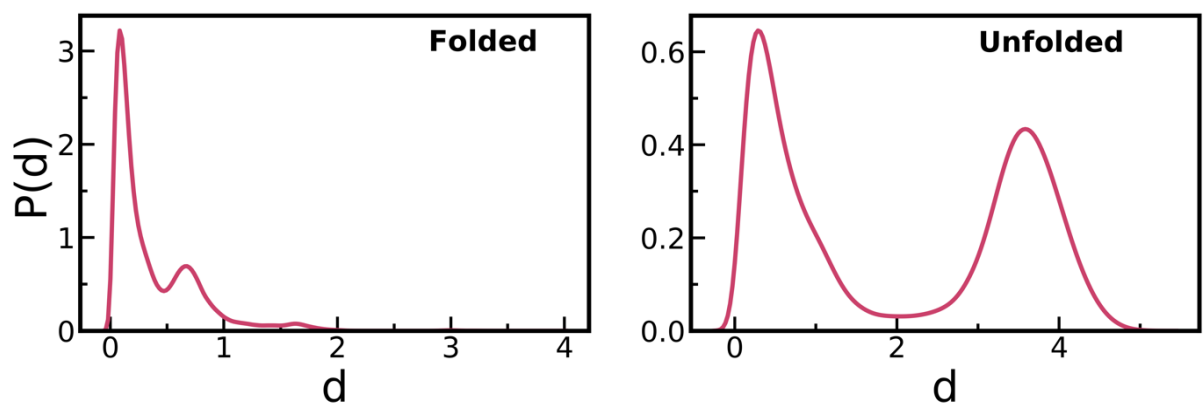

**Figure S4:** Distribution of pairwise distances between the projection points of the 1.85 μs long normal MD simulation trajectory frames on the TIC1 vs TIC2 plane in the folded and unfolded state basins for TC5b Trp-cage mutant.

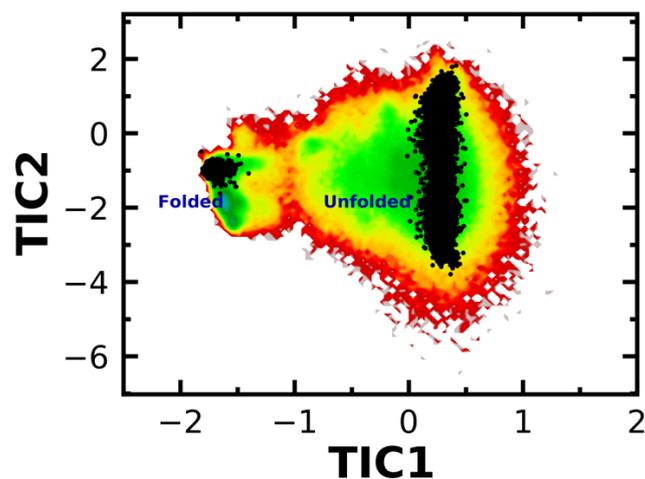

**Figure S5.** The projections (in black colour points) of the folded and unfolded state trajectories on the TICA eigenvectors trained using only the end-state simulation trajectories on the free energy surface derived using TICA eigenvectors trained on the full $168\,\mu s$ long Anton trajectory of Protein G. This shows that the TICA eigenvectors calculated from the end-state simulations capture the same slow process as the full length trajectory TICA model.

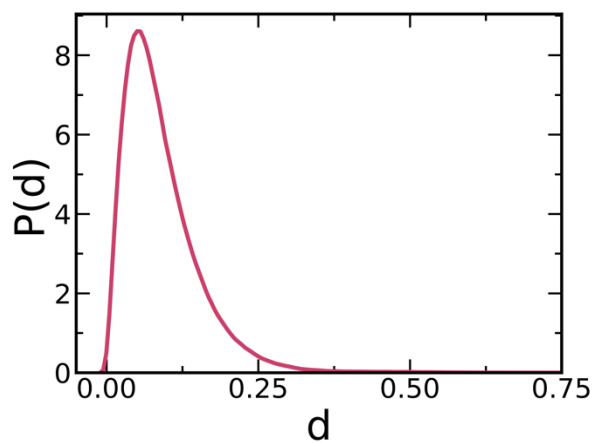

**Figure S6.** Distribution of pair-wise distances between the projection points of the $2\mu s$ long Anton trajectory frames on the TIC1 vs TIC2 plane in the folded state basin for Protein G.

References:

1) Lindorff-Larsen, Kresten, et al. "How fast-folding proteins fold." Science 334.6055 (2011): 517-520.